%% file: 0-elsevier.tex
\documentclass[5p,times]{cas-dc}

\usepackage[numbers]{natbib}
\usepackage{amssymb}
\usepackage{graphicx}%
\usepackage{multirow}%
\usepackage{amsmath,amssymb,amsfonts}%
\usepackage{amsthm}%
\usepackage[title]{appendix}%
\usepackage{textcomp}%
\usepackage{nicematrix}

\usepackage{booktabs}%
\usepackage{algorithm}%
\usepackage{algorithmicx}%
\usepackage{algpseudocode}%
\usepackage{listings}%

\usepackage{verbatim}
\usepackage{threeparttable} 

\usepackage{pifont}
\usepackage{array}
\usepackage{booktabs}

\newcommand{\github}{GitHub }
\newcommand{\mycheckmark}{\ding{51}}%
\newcommand{\cross}{\ding{53}\xspace}%
\usepackage{balance}
\usepackage{hyperref}
\usepackage{amsmath}
\usepackage[capitalize,noabbrev]{cleveref}
\usepackage{enumitem}
\usepackage{xspace}

\usepackage{url}
\usepackage{tablefootnote}
\usepackage{listings}
\usepackage{colortbl}
\usepackage{todonotes}
\usepackage{longtable}

\colorlet{punct}{red!60!black}
\definecolor{background}{HTML}{EEEEEE}
\definecolor{delim}{RGB}{20,105,176}
\colorlet{numb}{magenta!60!black}

\lstdefinelanguage{json}{
    basicstyle=\fontsize{8}{10}\ttfamily,
    numbers=left,
    numberstyle=\scriptsize,
    stepnumber=1,
    numbersep=4pt,
    showstringspaces=false,
    breaklines=true,
    frame=lines,
    backgroundcolor=\color{background},
    literate=
     *{:}{{{\color{punct}{:}}}}{1}
      {,}{{{\color{punct}{,}}}}{1}
      {\{}{{{\color{delim}{\{}}}}{1}
      {\}}{{{\color{delim}{\}}}}}{1}
      {[}{{{\color{delim}{[}}}}{1}
      {]}{{{\color{delim}{]}}}}{1},
}

\newcommand{\eg}{e.g.\xspace}

\usepackage{tikz}

\def\halfcheckmark{\tikz\draw[scale=0.4,fill=black](0,.35) -- (.25,0) -- (1,.7) -- (.25,.15) -- cycle (0.75,0.2) -- (0.77,0.2)  -- (0.4,0.7) -- cycle;}

\begin{document}
\let\WriteBookmarks\relax
\def\floatpagepagefraction{1}
\def\textpagefraction{.001}

\shorttitle{}    

\shortauthors{Costa, Barbosa, and Cunha}  

\title [mode = title]{A Framework for Supporting the Reproducibility of Computational Experiments in Multiple Scientific Domains}  



%

\author[inst1]{Lázaro Costa}
\ead{lazaro@fe.up.pt}

\author[inst2]{Susana Barbosa}
\ead{susana.a.barbosa@inesctec.pt}

\author[inst1]{Jácome Cunha}
\ead{jacome@fe.up.pt}

\affiliation[inst1]{organization={HASLab/INESC TEC, Faculty of Engineering, University of Porto},
            city={Porto},
            country={Portugal}}

\affiliation[inst2]{organization={INESC TEC},
            city={Porto},
            country={Portugal}}




\begin{abstract}
In recent years, the research community, but also the general public, has raised serious questions about the reproducibility and replicability of scientific work. Since many studies include some kind of computational work, these issues are also a technological challenge, not only in computer science, but also in most research domains.

Computational replicability and reproducibility are not easy to achieve due to the variety of computational environments that can be used. Indeed, it is challenging to recreate the same environment via the same frameworks, code, programming languages, dependencies, and so on.

We propose a framework, known as SciRep, that supports the configuration, execution, and packaging of computational experiments by defining their code, data, programming languages, dependencies, databases, and commands to be executed.
After the initial configuration, the experiments can be executed any number of times, always producing exactly the same results. Our approach allows the creation of a reproducibility package for experiments from multiple scientific fields, from medicine to computer science, which can be re-executed on any computer. The produced package acts as a capsule, holding absolutely everything necessary to re-execute the experiment.

To evaluate our framework, we compare it with three state-of-the-art tools and use it to reproduce 18 experiments extracted from published scientific articles. With our approach, we were able to execute 16 (89\%) of those experiments, while the others reached only 61\%, thus showing that our approach is effective. Moreover, all the experiments that were executed produced the results presented in the original publication. Thus, SciRep was able to reproduce 100\% of the experiments it could run.
\end{abstract}


\begin{highlights}
\item We provide a set of principles that reproducibility tools should follow
\item Researchers still face challenges when trying to reproduce scientific work
\item We propose a framework for reproducibility that covers multiple scientific domains
\item We provide a dataset of computational experiments for benchmarking
\item We provide a set of research artifacts for existing scientific publications



\end{highlights}



\maketitle

\section{Introduction} \label{sec:backend_intro}
\input{1-intro}

\section{Related Work} \label{sec:sota_tools}
\input{2-relatedWork}

\section{Challenges in Reproducibility} \label{sec:challengesInReproducibility}
\input{3-challenges}

\section{Researchers' Experience with Reproducibility Tools} \label{sec:survey}
\input{4-survey}

\section{Design and Implementation of a Reproducibility Framework} \label{sec:Methodology}
\input{5-methodology}

\section{Evaluation} \label{sec:experimental}
\input{6-experimental}

\section{Discussion} \label{sec:discussion}
\input{7-Discussion}

\section{Threats to Validity} \label{sec:threats}
\input{8-threats}

\section{Conclusion} \label{sec:backend_conclusion}
\input{9-conclusion}

\section*{Acknowledgements}
This work is financed by National Funds through the Portuguese funding agency, FCT - Fundação para a Ciência e a Tecnologia, within project UIDB/50014/2020. DOI 10.54499/UIDB/50014/2020

This research was supported by the doctoral Grant SFRH/BD/1513 66/2021 financed by the Portuguese Foundation for Science and Technology (FCT), and with funds from Portugal 2020, under MIT Portugal Program.


\balance


\bibliographystyle{elsarticle-num} 
 
\bibliography{bibliography}

\appendix

\section{API Rest} \label{app:api}
\input{appendixApi}




\end{document}

%% file: 1-intro.tex
Reproducibility is the ability to replicate the results of a study via the original methods, materials, and data, whereas replicability is the capacity to obtain the same results via new methods, materials, and data and/or conditions that are consistent with those of the original study. Both are paramount for science and are widely considered preconditions of scientific advancement \cite{Bush2020}.

In recent years, many concerns have been raised related to the reproducibility and replicability of today's scientific work. In particular, a significant amount of scientific studies are currently supported by some kind of computational experiment. Thus, reproducible and replicable research is now more than ever a technological challenge in many research domains.
Indeed, research in many fields (\eg, chemistry, climate change, biology) is supported, in several studies, by computational experiments~\cite{Ivie2018}.

Initiatives such as the ACM Badging Policies play a crucial role in promoting reproducibility in the Computer Science community by providing standardized assessment frameworks. While they encourage best practices and transparency, their impact remains limited as they do not enforce technical implementation or usability improvements.

Continuous Integration (CI) and Continuous Delivery (CD) automate code integration, dependency management, testing, and deployment. However, CI/CD alone does not fully automate reproducibility, as additional instructions are often required to reproduce an experiment accurately. Manual steps, such as specifying execution parameters, handling dependencies beyond package managers, and ensuring data availability, remain necessary to achieve complete reproducibility~\cite{Soares2022}.

To replicate and reproduce computational experiments, it is necessary to rebuild the original computing environment using the original code and its dependencies, the same data, and the same script-based workflow~\cite{Nust2021, Brunsdon2021}.
Computational reproducibility and replicability should be easily achievable since it is possible to specify detailed information about the computational environment being executed. This should enable the reproduction of the computation environment as in the original experiment. However, the complexity of today's software makes such reproduction difficult \cite{Ivie2018}.

Package management tools of each programming language (PL), such as pip\footnote{\url{https://pypi.org/project/pip/}} for Python, have integrated commands that can build a file with the dependencies needed for an experiment.
However, these files do not have all the information to enable computational reproducibility.

There are tools to aid researchers in making their computational experiments reproducible, such as 
Binder~\cite{Matthias2018binder}, Code Ocean~\cite{Codeocean}, 
Provenance-To-Use (PTU)~\cite{Pham2013PTU}, RenkuLab~\cite{ramakrishnan2023renku}, ReproZip ~\cite{Chirigati2020}, Sciunit~\cite{Ton2017SciUnits} or Whole Tale~\cite{Brinckman2019}.
However, these approaches are often limited to a set of PLs, do not allow the use of more than one PL per project, do not allow the use of databases, and do not allow researchers to specify which commands should be executed to achieve the experimental results.

%
%


In this paper, we present SciRep which can reproduce a wide range of computational experiments from multiple scientific fields, ranging from medicine to computer science. The goal of this framework is to cope with experiments with different requirements, from the PLs used to the databases.
With our work, we intend to answer the following research questions (RQs):\\

\noindent
\textbf{RQ1: What is the adoption of reproducibility tools among researchers?}\\

To answer this RQ, we designed and distributed an online survey.
Although the aforementioned approaches allow the reproducibility of some experiments, only 13 of the 102 researchers answering our survey acknowledged having used one of the computational reproducibility tools to save and reproduce computational experiments. We provide more details in \cref{sec:survey}.\\


\noindent
\textbf{RQ2: Can a single framework reproduce and replicate computational experiments from multiple scientific fields?}\\

To answer this RQ, we propose a novel framework that allows the creation, configuration, (re-)execution, packaging, and sharing of computational experiments, which also includes the validation of the reproduction and replication of the results via the original or a new dataset.
%
%
SciRep supports a variety of experiments from different scientific fields, including climate change, medical, and computer science. Moreover, it also supports artificial intelligence (AI) experiments, for which researchers are aware of the importance of making the experiment reproducible (\eg, by encouraging the explicit specification of seeds for pseudo-random number generators in non-deterministic analysis).
The proposed framework supports databases and automatically infers information from the experiment, such as the PLs used and dependencies, automatically creating the proper environment for execution. The framework allows the export of a research artifact that can be executed on any computer just by double-clicking a single file, making the experiment easily sharable and re-executable by anyone. Note that it is not necessary to have any knowledge about the underlying scientific work to re-execute it as it is performed fully automatically.

To evaluate our proposal, we selected a set of experiments included in previously published articles and recreated them via our framework. With this, we seek to answer the following RQ (\cref{sec:experimental}):\\

\noindent
\textbf{RQ3: What kinds of experiments can our framework support?}\\

To answer this RQ, we collected experiments from related works, articles published at major conferences, and others from Zenodo so that we could perform a variety of experiments. We collected 28 experiments from fields such as computer science, medicine, and climate change science. In general, we successfully reproduced the experiments, obtained the same published results, and created a research artifact that anyone can re-execute just by running a single file. Moreover, compared with other tools, SciRep was able to reproduce a broader range of experiments compared to other tools.

On the basis of these RQs and our answers, the main contributions of this work are as follows:

\begin{itemize}
    \item A set of principles that a reproducibility and replicability tool should follow (\cref{sec:challengesInReproducibility});
    \item An initial overview of the perspectives of researchers about reproducibility difficulties and tool usage (\cref{sec:survey});

    \item A framework for reproducibility and replicability, tackling the principles and the researchers' perspectives and coping with experiments from multiple scientific fields (\cref{sec:Methodology});

    \item A set of computational experiments to compare the efficiency of the new reproducibility tools (\cref{sec:experimental});

    \item A set of research artifacts for existing scientific publications via our approach (\cref{sec:experimental}).
\end{itemize}

In the following section, we present related work and contextualize our contribution.

%% file: 2-relatedWork.tex
In previous work, we conducted a literature review aimed at evaluating tools designed to assist researchers from various fields in addressing reproducibility challenges \cite{costa2024Rep}. 
We identified 18 tools: Binder~\cite{Matthias2018binder}, CARE~\cite{Janin2014}, Code, Data, Environment (CDE)~\cite{Philip2011}, Code Ocean~\cite{Codeocean}, Encapsulator~\cite{Pasquier2018}, FLINC~\cite{Ahmad2022Flinc}, PARROT~\cite{thain2005parrot}, PTU~\cite{Pham2013PTU}, Prune~\cite{Ivie2016Prune}, RenkuLab~\cite{ramakrishnan2023renku}, ReproZip~\cite{Chirigati2020}, reprozip-jupyter~\cite{reprojupyter}, Research\-Compendia~\cite{Stodden2015ResearchCompendia}, SciInc~\cite{Youngdahl2019SciInc}, Sciunit~\cite{Ton2017SciUnits}, SOLE~\cite{malik2014sole}, Umbrella~\cite{Meng2015Umbrella}, and Whole Tale~\cite{Brinckman2019}.
%
However, some tools are unavailable for download or are impossible to install or use. 
Therefore, only eight tools—Binder, Code Ocean, FLINC,
PTU, RenkuLab, ReproZip, Sciunit and Whole Tale—were used to reproduce three existing experiments across different domains.


We propose a set of characteristics to classify each reproducibility framework and separate these characteristics into four groups: \textit{i)} orthogonal characteristics, \textit{ii)} experimental configuration, \textit{iii)} reproducible procedure, and \textit{iv)} interoperability/development. These groups reflect the different stages of most computational experiments. \\

The first group ``orthogonal characteristics'' (C1-C3) includes:
\subsection*{C1: Supported Operating System}
Researchers often conduct their experiments on their preferred operating system (OS), and this preference should not hinder the reproducibility of an experiment. To address this, several tools have been developed to support all major OS, ensuring cross-platform compatibility. For example, Binder, Code Ocean, RenkuLab, and Whole Tale are notable examples that facilitate reproducibility across different OS platforms, thereby eliminating OS-specific constraints.

Conversely, some tools are limited to specific environments, which can be restrictive. FLINC, PTU, Sciunit, and ReproZip are confined to the Linux environment. This limitation may pose challenges for researchers who utilize other OS, potentially hindering their ability to reproduce experiments seamlessly.

Our work circumvents these limitations by operating as a web server, making it accessible to anyone with a web browser. Additionally, to re-execute the experiments, it relies solely on Docker\footnote{\url{https://www.docker.com/}}, which is compatible with all major OS. This ensures that our solution can be deployed and executed on any OS, thereby enhancing accessibility and reproducibility for researchers across different computing environments.

\subsection*{C2: Supported Programming Languages}
Researchers perform experiments using a variety of PLs. It is essential that the approach used can support such experiments. Tools such as Binder, Code Ocean, FLINC, RenkuLab, and Whole Tale support only a limited set of PLs and are not extendable, so it is impossible to support new PLs. However, ReproZip, Sciunit, and PTU support an unlimited set of PLs, as SciRep does.

\subsection*{C3: Supported Databases}
Experiments involving databases present a significant challenge for reproducibility, as the complexity of managing and replicating database environments adds an extra layer of difficulty. Consequently, most reproducibility tools do not support these types of experiments. However, ReproZip and our framework stand out as notable exceptions. Both tools are equipped to handle experiments incorporating databases, ensuring that the computational environment, including the database setup and data, can be accurately reproduced. This capability enables comprehensive reproducibility in computational research, where database-dependent experiments are increasingly common.

The ``orthogonal characteristics'' play a crucial role in broadly classifying the reproducibility platforms. \\

The second group, ``experimental configuration'' (C4-C6), includes characteristics to classify and define the experimental configuration procedure, such as:
\subsection*{C4: Automatically Detects Dependencies}
One of the main challenges in computational reproducibility is the ability of the framework to infer all the necessary dependencies and libraries to execute the code in the same way as the owner of the experiment.

FLINC, PTU, ReproZip, and Sciunit can automatically track and identify all the required information (\eg, dependencies, data, files, etc.) needed to execute the code. These approaches employ observed provenance capture to gather detailed provenance information and expose the underlying mechanisms of a script. 
On the other hand, Binder, Code Ocean, RenkuLab, and Whole Tale cannot detect all the necessary dependencies. 
SciRep can also automatically infer some dependencies.

\subsection*{C5: Automatically Installs Dependencies}
One of the main challenges in computational reproducibility is correctly installing the required libraries and other dependencies. 

Binder, FLINC, PTU, ReproZip, RenkuLab, Sciunit, and Whole Tale can automatically install the necessary dependencies.
However, Binder, RenkuLab, and Whole Tale can only install dependencies if they are in the specification file, which can be challenging for researchers who are less familiar with this type of technique. 
For Code Ocean, all dependencies must be added manually by the researcher to be installed. 

Our framework can also automatically install all the dependencies in an automated way.

\subsection*{C6: Automatically Configures the Execution of Experiments}
Ensuring computational reproducibility requires making all relevant information available, including the specific values of the parameters used during the execution of experiments.
Several frameworks support this feature by enabling researchers to share parameter spaces with their peers. Notable examples include Code Ocean, PTU, ReproZip, and Sciunit. These tools facilitate transparency and reproducibility by allowing researchers to document and share the parameters explored in their studies.

Similarly, our framework empowers researchers to define and share the parameters of each experiment. By doing so, it ensures that others can accurately replicate and build upon their work, fostering a collaborative and transparent research environment.


The emphasis on the ``experimental setup'' criteria is crucial, as it significantly impacts the efficiency and reliability of reproducibility tools, guaranteeing a simplified and standardized approach. \\

The third group, ``reproducible procedure'' (C7) includes the characteristic of the procedure used to reproduce the experiment:

\subsection*{C7: Research Artifact}
The integrity and reproducibility of research findings heavily depend on the availability and accessibility of essential research artifacts, such as code, datasets, and dependencies, that form the foundation of scientific discoveries. It is imperative that these artifacts are well-documented, easily accessible, and capable of being replicated across diverse environments. 
While tools such as Binder and RenkuLab do not facilitate the creation of research artifacts, others, such as Code Ocean, FLINC, PTU, ReproZip, Sciunit, and Whole Tale, excel in building and managing these critical components of experiments. These tools are pivotal in enabling researchers to create, share, and reproduce research artifacts effectively.

Our work allows us to easily create a research artifact that can be executed on any computer, requiring only one dependency, namely, Docker. Docker is a platform designed to help developers build, share, and run modern applications, and is the platform most commonly used for this purpose~\cite{rad2017introduction}. Given its wide acceptance, it is well-maintained and easy to obtain and install. However, it has been designed to be used by professional software developers and is not usable by typical researchers. Our framework builds on Docker but abstracts its complex usage by providing a set of features that are at the same level of abstraction that researchers are used to accustomed thinking.

The ``reproducible procedure'' criteria is essential to characterize the creation of the reproducibility package.\\

The fourth group, ``Cross-Platform Support and Devolvement'' (C8-C10), includes characteristics such as:

\subsection*{C8: Ensuring the Reproducibility of the Results}
Ensuring the reproducibility of the results reported in publications can be challenging and time-consuming. Researchers need to replicate the experiment carefully, read the results section, and compare their own findings with those documented in the publication. To simplify this process, it is beneficial for researchers to register the expected results of an experiment in advance. 
This allows automatic comparisons during each run, making it easier to verify reproducibility. However, the current approaches 
often lack this crucial functionality. Our approach addresses this gap by incorporating features that allow researchers to register the expected outcomes and automatically compare them with the actual results during each execution. This not only simplifies the validation process but also enhances the reliability and credibility of the published results.

\subsection*{C9: Reproducibility Across Operating Systems}
Interoperability across different OS is a critical factor in ensuring the widespread usability and accessibility of research tools. When a tool is interoperable in relation to OS, it can function seamlessly across various OS platforms, such as Windows, macOS, and Linux. This broad compatibility is essential for facilitating collaboration among researchers who may use different systems on the basis of their preferences, institutional policies, or specific requirements of their work.

Tools such as Binder, Code Ocean, RenkuLab, and Whole Tale, as well as our tool, exemplify interoperability, supporting all major OS and facilitating cross-platform reproducibility of experiments. However, tools such as Flinc, PTU, ReproZip, and Sciunit are limited to the Linux environment, which can restrict their usability for researchers using other OS. Our approach overcomes these limitations by operating as a web server, being accessible through any web browser, and relying solely on Docker, ensuring compatibility with all major OS. This design enhances accessibility and reproducibility for researchers across diverse computing environments.

\subsection*{C10: Developed Environment}
In the domain of tool development, two primary environments often contribute significantly to creating innovative and impactful tools: the academic environment (Acad.) and the enterprise environment (Ent.). Each setting fosters a unique approach to development driven by different goals, methodologies, and constraints. 

Academic tools such as Binder, FLINC, PTU, RenkuLab, ReproZip, Sciunit, and Whole Tale are developed in research institutions or universities, emphasizing experimental, cutting-edge technology and theoretical advancements. On the other hand, enterprise tools, such as Code Ocean, are created within corporate settings, with a focus on practical application, market viability, and user needs.

The ``Cross-Platform Support and Development'' criteria comprise key characteristics vital for the approaches' sustainability and compatibility.

\subsection*{Summary}

\begin{table*}[!tbh]
\centering
\rowcolors{2}{gray!20}{white}
  \centering
\caption{Comparison of the characteristics of the reproducibility tools}
    \begin{tabular}    
    {>{\centering}m{1.3cm}
    >{\centering}m{1.5cm}
    >{\centering}m{2.2cm}
    >{\centering}m{1cm}
    >{\centering}m{1.1cm} 
    >{\centering}m{2.3cm}
    >{\centering}m{1.5cm}
    >{\centering}m{1cm}
    >{\centering\arraybackslash}p{1cm} }
    \rowcolor{gray!50}
    \toprule
    
        Charact. & RenkuLab & Whole Tale & Reprozip & Sciunit & Code Ocean & Binder & PTU & Our \\ 
        \midrule
        C1 & All & All & Linux & Linux & All & All & Linux & All \\ 
        C2 & R, Python, Julia, MATLAB & R, IPython, IRKernel, Julia, MATLAB & Unlim. PLs & Unlim. PLs & Python, R, Java, C, C++, Scala, Lua, MATLAB & IPython, IR-Kernel, IJulia & Unlim. PLs & Unlim. PLs \\ 
         C3 & \cross & \cross & \mycheckmark & \cross & \cross & \cross & \cross & \mycheckmark\\
        C4 & \cross & \cross & \mycheckmark & \mycheckmark & \cross & \cross & \mycheckmark & \halfcheckmark\\ 
        C5 & \mycheckmark & \mycheckmark & \mycheckmark & \mycheckmark & \cross & \mycheckmark & \mycheckmark& \mycheckmark \\ 
        C6 & \cross & \cross & \mycheckmark & \mycheckmark & \mycheckmark & \cross & \mycheckmark & \mycheckmark\\ 
        C7 & \cross & \mycheckmark & \mycheckmark & \mycheckmark & \mycheckmark & \cross & \mycheckmark & \mycheckmark\\ 
        C8 & \cross & \cross & \cross & \cross & \cross & \cross & \cross & \mycheckmark\\ 
        C9 & \mycheckmark & \mycheckmark & \mycheckmark & \cross & \mycheckmark & \mycheckmark & \cross& \mycheckmark \\ 
        C10 & Acad. & Acad. & Acad. & Acad. & Ent. & Acad. & Acad. & Acad. \\
        \bottomrule
    \end{tabular}
      \label{tab:toolComparison}%
\end{table*}

\cref{tab:toolComparison} synthesizes the characteristics of the approaches that we have just described. 
For each characteristic (first column), each approach (following columns, ours being the last one) is marked accordingly: if the approach fully supports the characteristic, it is marked with \mycheckmark; if the approach does not support the characteristic, it is marked with \cross; if the approach partially supports the characteristic, it is marked with \halfcheckmark. This assessment has been performed in previous work \cite{costa2024Rep}.

We conclude that these approaches 
have some limitations and cannot support the entire reproducible process of a computational experiment. Our approach is the one that covers more characteristics, followed by ReproZip as only these approaches support databases. However, ReproZip is limited to Linux and does not verify whether the result of each execution is the same as before; that is, if the experiment has been truly reproduced, it cannot cope with notebooks, which are often used by researchers.

%% file: 3-challenges.tex
In this section, we summarize several challenges identified in the literature, relating our approach to them. Moreover, we use these challenges to define a set of principles that a tool aiming to aid in reproducibility should follow to address such challenges.

\subsection{Challenges}\label{sec:challenges}

The digital transformation of science has made new publication channels for research data available. These publications come from specialized data repositories, commercial application providers, or even traditional self-hosted web server downloads.
Some examples of these frameworks are 
CKAN\footnote{\url{https://ckan.org/}}, DSpace\footnote{\url{https://dspace.lyrasis.org/}}, EUDAT\footnote{\url{https://www.eudat.eu/}}, 
GitHub\footnote{\url{https://github.com/}}, 
Harvard Dataverse\footnote{\url{https://dataverse.harvard.edu/}}, and Zenodo\footnote{\url{https://zenodo.org/}}.
Currently, these frameworks provide ways to upload and publish research datasets with an institutional, disciplinary focus, or general purpose. These frameworks also need to address the needs of different stakeholders in a variety of usage scenarios~\cite{Langer2019, kim2018internet}.

Although these frameworks help spread science, they are not enough to make it reproducible.
Researchers need to expose not only their results but also other artifacts, such as code and data, and explain how to use them to achieve the published results. However, these frameworks do not provide means to make the code easily reproducible.

On the basis of previous work, we summarize a set of challenges a researcher needs to consider when intending to make an experiment easily reproducible:

\begin{description}[leftmargin=8pt]
  \item[\textbf{Challenge 1: Software}]
    In the research process, software greatly impacts the reproducibility of experiments~\cite{Sarah2017}.

    When someone tries to reproduce an experiment but uses the wrong software (\eg, dependencies and their versions), the experiment may produce different results from the original one. It is known as \textit{code rot}~\cite{boettiger2015}. For example, some newer versions of software may claim to be backward compatible; however, they may have unintended differences that can affect reproducibility. 
    Thus, all the information related to the software should be identified to ensure a consistent behavior~\cite{Ivie2018}.
    Another possible scenario is related to the \textit{dependency hell}, a term that refers to a common issue in software development where multiple packages share a dependency. 
    As only one version of a dependency is allowed in a project's environment, resolving these conflicts and finding a compatible solution can be challenging~\cite{Pham2015}.
    Another potential problem is when the libraries are unavailable. 
    This may be challenging or even impossible, hindering reproducibility.  

    Software is a crucial tool in assisting researchers with the reproducibility process. However, finding complete and easy-to-use solutions remains a challenge~\cite{Freire2012_b, Zuduo2021}, as we have also detailed in the previous section.

    The framework we propose addresses this challenge as it provides mechanisms to define experiments from multiple scientific domains with any number of PLs, dependencies, datasets, and so on. Moreover, our platform creates a capsule with all the required software to re-create an experiment in a straightforward manner.

   \item[\textbf{Challenge 2: Command}]
    It is crucial to make the commands available to execute the code, along with the file to execute and all the values of the parameters needed to reproduce the experiment.
    Therefore, using parametrization is an important method for extension,
    as it allows researchers to easily extend the approach, explore the parameter space, and validate the results~\cite {aslst2002workflow,liu2015survey, Ivie2018}.
  
    Furthermore, the expected result should be available for comparison because there is no guarantee that the same command applied by a different researcher on a different machine will obtain the same result~\cite{Ivie2018}. 

    We also address this challenge, as SciRep provides mechanisms to define the commands, as well as their parameters, that are necessary to execute an experiment. Moreover, the framework allows the definition of the expected outputs to be compared with future executions.
  
  \item[\textbf{Challenge 3: Data}]
    Making all the data available is another important element to allow reproducibility~\cite{David2019}.
    Many researchers use initial data to start their experiments and then produce results on the basis of those data. Sometimes, these data are stored in places that are not nearby, such as remote servers. Therefore, there can be security or privacy concerns when accessing or using the data.

    One way to improve reproducibility without risking sensitive data is to share details about the data, such as how it can be accessed or used, instead of sharing the actual data itself. In this way, others can understand and replicate the research without having direct access to sensitive information~\cite{Ivie2018}.

    Furthermore, researchers intend to find and reuse existing data, but the current approaches do not include sufficient metadata (\eg, title, authors, version, PLs, software description, DOI, and execution requirements) to find and reuse data~\cite{Stodden2016}.

    Our platform allows the creation of a capsule with all the code and the necessary data to execute it. However, currently, we provide limited support for metadata. Thus, we address this challenge practically but intend to work on it in future work. 
  
   \item[\textbf{Challenge 4: Hardware}] 
    Enabling reproducibility using the same performance, scalability, and efficiency is another challenge since the same hardware should be used (\eg, the same computing power, such as RAM, CPU, GPU, and storage). 
    However, in this work, we do not consider experiments that may have hardware requirements.
  \end{description}

\subsection{Principles}\label{sec:principles}

From challenges 1, 2, and 3 discussed above, we identify a set of principles that a reproducibility tool should follow. We also introduce principles that are orthogonal to the challenges.

\begin{description}[leftmargin=1em]
\item[From Challenge 1]  \ \\  \vspace{-.7cm}
    \begin{description}[leftmargin=0em]
       \item[\textbf{P1}] The reproducibility framework should allow researchers to define the PLs used in the experiment, the compilers, the dependencies, and the libraries, as well as the correct version of each.

       \item[\textbf{P2}] It should allow saving all the dependencies needed to execute the experiment together with the code.
    \end{description}

\item[From Challenge 2] \ \\ \vspace{-.7cm}
    \begin{description}[leftmargin=0em]
        \item[\textbf{P3}] The framework should support the configuration of the commands for the correct execution of the experiment.

        \item[\textbf{P4}] It should allow the validation of the computational reproducibility using the results of previously performed experiments as a means of comparison.
    \end{description}

\item[From Challenge 3] \ \\ \vspace{-.7cm}
    \begin{description}[leftmargin=0em]
        \item[\textbf{P5}] Such a framework should provide mechanisms that allow researchers to manage the data of the experiment.
    \end{description}

\item[Orthogonal principles] \ \\ \vspace{-.7cm}
    \begin{description}[leftmargin=0em]
        \item[\textbf{P6}] It should be able to address the diversity of PLs (and their dependencies) used by researchers.

        \item[\textbf{P7}] It should support all types of experiments, from the execution of simple scripts to more complex experiments (\eg, involving databases or several PLs).
    
        \item[\textbf{P8}] The framework should support researchers who are not computer science experts.
        
        \item[\textbf{P9}] It should easily allow sharing the experience.

        \item[\textbf{P10}] It should facilitate the re-execution of the experiment even for users outside the research domain.
    \end{description}

\end{description}

In \cref{sec:Methodology}, we introduce our platform and how it includes these principles.


In the following section, we present a small survey about the views of researchers on reproducibility, gathering further evidence to support the need for the platform we now propose.

%% file: 4-survey.tex
Previous studies have investigated the willingness of researchers to share code and data, the extent to which research artifacts are available, and the success rates of reproducing published experiments~\cite{Stodden2010, Prabhu2011, Antunes2024, Ivie2018}. However, less attention has been given to the specific challenges faced by researchers in attempting to reproduce computational experiments, making it difficult to pinpoint key barriers and areas for improvement.

Thus, we intend to find further motivation for our work by directly investigating the experiences of researchers with reproducibility. Instead of focusing solely on sharing or the availability of code and data, we devised a preliminary survey that examines the practical difficulties faced by those trying to reproduce experiments. 

Thus, we performed an online survey\footnote{Available at \url{https://forms.gle/PMHck2dmH7Qms8Ft9}.} to gather information related to the familiarity of researchers with the concept, the challenges, requirements, tools, and the main difficulties of reproducibility of computational experiments. 

The results of our survey offer a preliminary insight into the current challenges of reproducibility, serving primarily as a motivational starting point for our work.


We made this survey available to all the authors of the 2022 edition of the International Conference on Software Engineering since we used some of these works in our evaluation (see \cref{sec:experimental}). From the 700 contacted authors, we got 38 responses. Moreover, we spread the same survey on social media platforms (Twitter, Facebook, and LinkedIn) through public posts and also distributed the survey through our contacts via email, for which we obtained 64 responses.

Our sample is composed mostly of Ph.D. students (46\%), followed by faculty members (29\%) and researchers (12\%). The majority are from Portugal (47\%), the USA (13\%), or China (12\%). Furthermore, 93\% of the respondents had more than two years of research experience.

The ``computational reproducibility'' term is familiar to 81\% of the surveyed participants, and 83\% tried to reproduce experiments performed by others. However, most of them, 66\%, had difficulties doing so, meaning that only a few researchers did not find it difficult to reproduce others' work.

The main challenges in computational reproducibility found in this survey are ``missing information to execute the experiment'' (79\%), ``data not available'' (63\%), ``code not available'' (63\%), and ``installation not successful'' (50\%).

Only 21\% of the surveyed individuals were familiar with a reproducibility framework, with Binder being the most recognized (18\%), followed by ReproZip (4\%) and RenkuLab (2\%).
Only 13\% of the participants reproduce their experiments using Docker or virtual machines (11\%), while 8\% share their code on Zenodo (5\%) and 3\% on GitHub.

Finally, we asked what requirements were considered necessary to allow computational reproducibility. The answers were ``make available the data'' (91\%), ``make available the code'' (88\%), ``make available the description of all the dependencies used and its versions'' (76\%), and ``make available the PL used and its version'' (56\%).

\subsection{Answering RQ1}
The survey we conducted, despite its limited scope, offers valuable insights into the current landscape of computational reproducibility from the perspective of researchers.
Our survey results indicate that the adoption of reproducibility tools among researchers is very limited. Only 21\% of respondents are aware of frameworks such as Binder, ReproZip, and RenkuLab, and even fewer use tools such as Docker (13\%) or publish their code on Zenodo (5\%) or \github (3\%). Despite the widespread recognition of the importance of computational reproducibility, practical implementation is hindered by challenges such as missing information (79\%), unavailable data (63\%), inaccessible code (63\%), and installation not successful (50\%). These findings highlight a significant gap between awareness and the actual use of reproducibility tools, underscoring the need for better tools, dissemination, training, and incentives to encourage the adoption of these tools among researchers.

We can conclude that although there are already some proposals to aid in reproducing computational experiments (see \cref{sec:sota_tools}), the researchers who answered this survey still do not use them at scale, which also motivated our own work.

%% file: 5-methodology.tex
In this section, we present our design and implementation of a framework, which we term SciRep, that provides support for the creation and execution of computational experiments. SciRep is based on the principles defined in \cref{sec:challengesInReproducibility} and the data collected from the researchers described in \cref{sec:survey}.
The focus of our work is on the software environment, and we assume that the hardware does not influence the outcome of the experiments (although we understand that in some kinds of experiments, it actually happens, we will not consider such cases).
In this work, we do not provide a user interface, which we plan to design and build in future work \cite{costa2022platform,costa2024vlhccGC}. Our main goal is to provide a framework that can cope with a wide range of experiments.

We present the features we envision for the framework (\cref{sec:backendFeatures}), its architecture (\cref{sec:backendArchitecture}), our current implementation (\cref{sec:implementation}), the API it provides (\cref{sec:api}), and an execution example to illustrate its use (\cref{sec:examples}).

\subsection{Framework Features}\label{sec:backendFeatures}
\begin{figure*}[!thb]
\centerline{\includegraphics[width=1\textwidth] {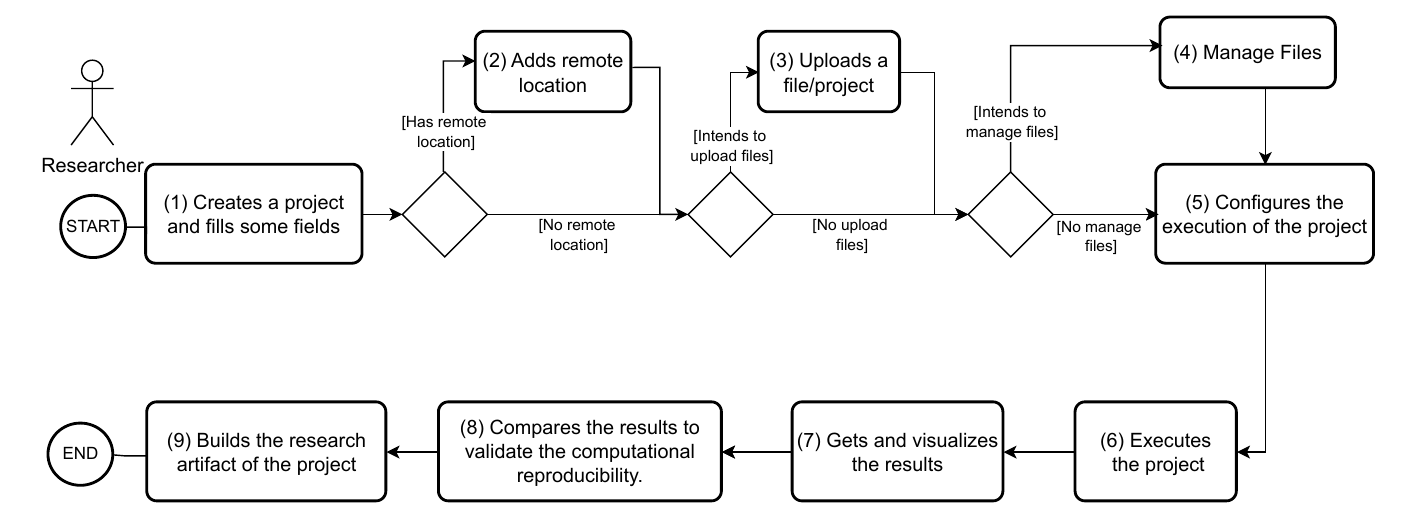}}

\caption{An activity diagram presenting the process a researcher should follow to create a research artifact via SciRep}
\label{fig:activityDiagram}
\end{figure*}

Ensuring reproducibility in computational research is challenging due to the complexity of software environments, dependencies, execution parameters, and results validation. SciRep simplifies this process by providing researchers with a structured workflow for setting up, executing, and sharing computational experiments. In the following, we outline its core features, grouped into three key phases: Experiment Setup (\cref{sec:exp}), Execution and Validation (\cref{sec:exec}), and Artifact Creation and Sharing (\cref{sec:arti}).
These phases and their steps are illustrated in \cref{fig:activityDiagram}. 

\subsubsection{Experiment Setup and Configuration}\label{sec:exp}
Reproducible research begins with a well-defined project structure. Our framework allows researchers to initialize a new project by specifying essential metadata such as the project name, description, and experiment type (Step 1 in the activity diagram present in \cref{fig:activityDiagram}). The type of experiment is particularly important, as it determines additional required configurations. For example, AI experiments often involve pseudo-randomness, requiring explicit seed values to ensure consistent outputs.

To simplify setup, researchers can import existing projects from external repositories, such as GitHub, ensuring the reproducibility of previous works (Step 2). If additional files are required, researchers can upload experiment-related files, including datasets, scripts, and dependencies (Step 3). These files can be organized, renamed, and managed within the framework itself (Step 4), providing flexibility in handling research data. This step realizes principle P5.

The inference mechanism in our framework automates the identification of PLs and their dependencies, minimizing manual setup efforts (Step 5). This process consists of two stages: PL identification and dependency resolution.
In the first stage, the framework analyzes file extensions (e.g., \texttt{.py} for Python, \texttt{.cpp} for C++, \texttt{.java} for Java) and project structure conventions (e.g., \texttt{requirements.txt}, \texttt{pom.xml}, \texttt{Makefile}) to determine the languages used. 

In the second stage, the framework scans source code files for import statements, dependency declarations, and package manager configurations. For example, it detects import statements in Python files and dependencies in \texttt{pom.xml} files used by Maven in many Java projects.

Once dependencies are identified, the framework automatically generates a Dockerfile, incorporating the necessary system packages and dependencies. The generated Dockerfile provides a fully configured execution environment, ensuring that all required dependencies are installed and locked to specific versions. This prevents inconsistencies caused by software updates or system differences. This step realizes principles P1, P2, P6, and P7.

Currently, our framework supports all the PLs used in the evaluated experiments, including Java, C++, Python, R, and Unix Shell. Additionally, the system is designed to be extensible, allowing the integration of additional PLs as needed.

The framework also supports database-driven experiments by allowing users to configure database environments directly. Currently, we support SQL (MySQL, PostgreSQL, SQLite) and NoSQL (MongoDB) databases. Although SQLite provides a lightweight, file-based solution that does not require a separate service, more advanced experiments may require a dedicated database system. For example, experiment E4, used in our study (\cref{tab:dataset}), relies on a database for structured data storage and retrieval, making the availability of a full database service necessary. Our framework ensures that the appropriate database engine is included in the reproducibility package, along with the required configurations and dataset.

\subsubsection{Automated Execution and Validation}\label{sec:exec}
Once an experiment is configured, its execution must be reproduced under identical conditions. Researchers can define the commands needed to build and run the experiment (Step 6), ensuring that all execution steps are explicitly documented. This step realizes principle P3. 

To ensure reproducibility, the framework captures and stores execution outputs (Step 7). These outputs can be stored as console logs or saved as structured files, depending on the experiment’s requirements. A key challenge in reproducibility is verifying that each execution produces consistent results. Our framework includes a validation mechanism (Step 8) that checks if the outputs remain unchanged across different runs—If deviations occur, the system flags inconsistencies, prompting researchers to investigate potential issues, such as missing dependencies or changes in libraries.
This step realizes principle P4. 

The framework also supports computational replicability by allowing researchers to re-run experiments with different datasets, allowing them to validate whether results hold under new conditions. This feature is particularly useful for AI experiments, where models need to be tested against unseen data to assess their generalizability.
However, for complex experiments, adjustments may be required in pre-processing steps, data format handling, or model parameter configurations.
These modifications can introduce new dependencies or unforeseen issues that were not present in the original package.
Currently, SciRep only supports direct dataset exchange.

\subsubsection{Research Artifact Creation and Sharing}\label{sec:arti}
Reproducibility is not just about running experiments—it is also about making them accessible and easy to share. Our framework allows researchers to generate a research artifact that encapsulates the entire experiment (Step 9). This step realizes principles P9 and P10. 
The artifact includes all project files (code, datasets, configurations), a fully configured computational environment (via Docker), and the necessary execution commands.
This artifact ensures that anyone can reproduce the experiment without needing to manually install dependencies. The only necessary application is Docker, which is the most commonly used containerization tool \cite{dockeribm,dockeroracle, Amit2020}. Following best practices for writing Dockerfiles ensures reproducibility in data science workflows~\cite{Nust2020}.

An essential aspect of reproducibility is data provenance—tracking the origin, transformations, and usage of data throughout the research process. Although our framework ensures that computational environments remain intact, it does not explicitly manage data findability, accessibility, interoperability, and reusability (FAIR principles)~\cite{Wilkinson2016}. Once an experiment is packaged into a containerized research artifact, the data within it is not independently accessible or searchable, which limits direct alignment with FAIR principles. 
To enhance FAIR compliance, researchers can integrate these artifacts with external repositories, persistent identifiers (\eg, Digital Object Identifiers (DOIs)\footnote{\url{https://www.doi.org}}), and metadata-rich platforms that facilitate findability. Future work could explore automated metadata extraction and linkage to FAIR-compliant repositories.


\subsection{Architecture}\label{sec:backendArchitecture}
\begin{figure}[!htb]
\centerline{\includegraphics[width=\columnwidth]{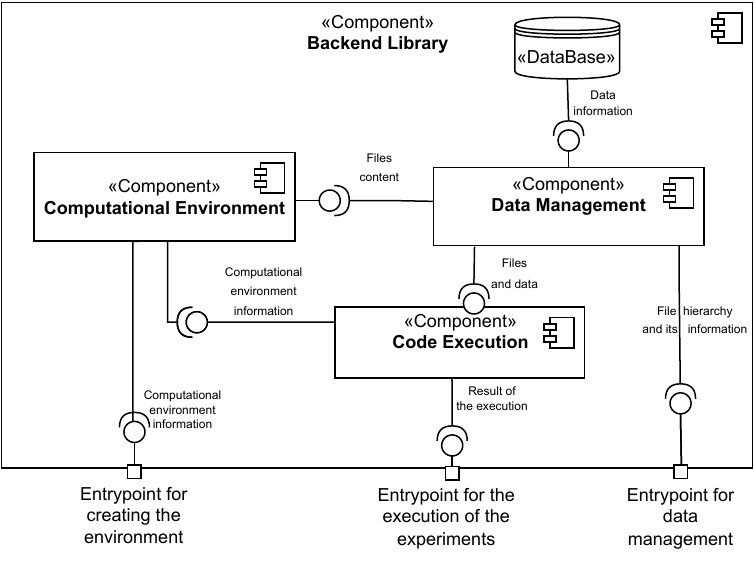}}

\caption{The components' diagram of our framework}
\label{fig:components-diagram}
\end{figure}

The proposed framework comprises three core components, namely, ``Data Management'', ``Computational Environment'', and ``Code Execution'', along with a dedicated database for efficient information storage. \cref{fig:components-diagram} illustrates the component diagram, followed by a detailed explanation of each component.

\subsubsection{Data Management}\label{sec:data}
This component is responsible for managing all the framework data, such as adding, editing, and deleting files and folders.
The Data Management component saves all the information of each file/folder in the database and manages it in the file system.

\subsubsection{Computational Environment}\label{sec:environment}
This component builds an environment capable of addressing the diversity of development stacks  (\eg, a variety of PLs, dependencies, and, if needed, databases) used by researchers and making that information available to the Code Execution component (\cref{sec:code}), which executes the experiment.

The diversity of available computing environments is a challenge in general software development and is commonly approached using \textit{containers}~\cite{clyburne2019computational, Amit2020}, a technology that allows automatically building an environment on the basis of a specification.
A specification file for a container is a document that outlines the configuration and requirements of a containerized application or service. It describes the resources, environment, dependencies, and other necessary information for the container to run consistently across different environments~\cite{clyburne2019computational, Amit2020}. 
By using the exact specification, each member of a team works in the exact same environment as all the others. 

Our framework adopts this principle to enable researchers to reproduce their experiments systematically, as the Computational Environment component is based on containerization.
The most commonly used containerization technology is Docker~\cite{Amit2020,dockeribm,dockeroracle}, which is the supporting tool our framework utilizes.

Notably, researchers do not need to have prior knowledge of containers or Docker. A key objective of our framework is to abstract away such complexities, allowing researchers across various domains to build research artifacts without requiring expertise in containerization technologies. Although our framework currently does not provide a user interface, we plan to develop one in future work to simplify artifact creation further \cite{costa2022platform,costa2024vlhccGC}.
For now, our focus remains on building a framework that can support the wide range of experiments published across scientific disciplines, from data science to medicine and the social sciences.


\subsubsection{Code Execution}\label{sec:code}
This component executes the experiments, validates the reproducibility or replicability of the results, and builds a research artifact with all the information needed to re-execute the experience in the same way.
This artifact is a Docker image, that is, it is a single file containing everything necessary to execute the experiment, from the (mini) OS, to the PLs' compilers, to the experiment's code and data. This single file is a capsule that does not require any further installation, update, or anything else. Thus, even if compilers, libraries, and other dependencies change, this capsule will not change and will execute exactly as when it was created.

\subsection{Implementation}\label{sec:implementation}
In this section, we describe our implementation in detail.

The Data Management component is implemented in Node.js\footnote{Available at \url{https://nodejs.org/en}.}, and we use the Express\footnote{Available at \url{https://expressjs.com/}.} framework for communication. 

The Computational Environment component is implemented in Python. We use the Flask\footnote{Available at \url{https://flask.palletsprojects.com/}} framework for communication and Docker Hub\footnote{Docker Hub is the world's largest container image repository containing open source projects (\url{https://hub.docker.com/}).} as containerization technology. 

Our approach is extensible and can build a variety of computing environments needed for researchers to execute their experiments. Our approach already supports multiple PLs, namely C++, Perl, R, Java (maven), Python, Unix Shell, and Jupyter Notebook. 
However, the extension of the PLs must be performed by someone with experience in containers.

Our approach supports databases and allows researchers to configure their setup. The databases currently supported are MongoDB, SQLite, MySQL, and PostgreSQL.
As with PLs, it allows the addition of other database techniques.

The implementation is available as open-source software at \url{https://github.com/lazarocosta/Framework}.

\subsection{An API for Reproducibility and Replicability}\label{sec:api}
In this work, we have implemented all the data management functionalities and all the procedures for building and managing the execution of experiments in our framework. We describe them in detail below. 

SciRep exposes an API that can be used by others to create their own reproducibility and replicability framework by providing a user interface. For example, researchers in the medical field could provide an interface that is more suitable for doctors, whereas computer scientists could provide a command-line user interface, which many researchers in the field are able to use (and often prefer).

Our framework provides an API to support the reproducibility and replicability of computational experiments. This API allows users to create projects, upload experimental files, build computational environments, execute experiments, and generate research artifacts.

We provide a small but powerful API. While this approach is not intended to be used by researchers directly, it provides a tool for others to build their own reproducibility environment for researchers, which can then be tailored for specific scientific domains. Indeed, we are currently working on an interface that uses such an API \cite{costa2024vlhccGC,costa2022platform}. 

A summary of the key API functionalities is presented below, while a detailed description of each endpoint is available in \cref{app:api}.

\begin{description}
    \item[\texttt{/project (POST)}] Create a new project and fill in the necessary fields, such as the project's name and description.
    
    \item[\texttt{\textless projectuuid\textgreater/uploadfile (POST)}] Upload a new file to the project.
    
    \item[\texttt{\textless projectuuid\textgreater/uploadproject (POST)}] Upload a complete pro- ject via a zip file.
    
    \item[\texttt{\textless projectuuid\textgreater/uploadgitproject (POST)}] Use a remote location to copy the content of a pre-existing project from GitHub.

    \item[\texttt{/\textless projectuuid\textgreater/build-docker-image (POST)}] Create a Docker file with all the necessary information (PL, dependencies, data, etc.) to run the project. After processing this file, the framework creates a Docker image with all the information described.

   
    \item[\texttt{/\textless projectuuid\textgreater/run-container (POST)}] Start a new execution of the project (new container) using the parameters received.

    \item[\texttt{/\textless projectuuid\textgreater/package (GET)}] Return the research artifact of the project, allowing the re-execution of the same experiment in the same way (environment and parameters) on any other computer.
    
\end{description}


\subsection{Execution Example} \label{sec:examples}
In this section, we provide examples of the execution of our framework.
We select two experiments (E3 and E7) from our evaluation (see \cref{sec:experimental}), and for each one, we start by creating and populating two new projects with the corresponding code (\cref{sec:ex_create}). We then build the necessary environment (\cref{sec:ex_env}) and execute the experiments (\cref{subsection:runExperiment}). Finally, we create a research artifact (\cref{subsection:reproducible_package}).

\subsubsection{Create a project and upload its files}\label{sec:ex_create}
During this task, we use the Data Management component described in \cref{sec:data}. 
In particular, we use the endpoint \texttt{/project} to create a project and fill in two parameters (the project name and description).

We then use the endpoint \texttt{\textless projectuuid\textgreater/uploadproject} to upload files to the project; as the parameter, we use the zip file of the project. 

At the end of these two API calls, a new project with the necessary files is created in the framework.

\subsubsection{Build the environment for each experiment}\label{sec:ex_env}
For this task, we used the Computational Environment component described in \cref{sec:environment} to build the environment for each experiment.
We exemplify the building process of two environments for two distinct experiments (E3 and E7). 

%
We start with E3, which uses C++ 
as the PL.
The documentation of E3 indicates that to build the experiment, it is necessary to run the two following commands: \textit{i)} \texttt{g++ -O3 ./src/bbfs\_node.cpp -o out}; and \textit{ii)} \texttt{g++ -O3 ./src/bbfs\_edge.cpp -o out}. 
%
Thus, we use the endpoint \texttt{/\textless projectuuid\textgreater/build-docker-image} without parameters because our approach infers the PLs (C++) and the dependencies to build the project.
After receiving the request, the server generates a dockerfile, builds a Docker image and returns an identifier (ID) that represents that Docker image.

In Listing~\ref{lst:dockerfile4}, we can see the dockerfile generated for E3. The directory ``files'' contains all the experiments' files. 
The first line indicates the use of the Ubuntu OS, followed by its update on line 2. Lines 3 and 4 install and update a C++ compiler. Lines 5 and 6 ensure that all the experimental data are available for use. Finally, the last two lines contain the commands to build the experiment. Thus, when this image is executed, it stops only when the experiment is built and ready to be executed.

\begin{lstlisting}[language=json,firstnumber=1, caption={Dockerfile of E3},captionpos=b, label={lst:dockerfile4} ]
FROM ubuntu:20.04
RUN  apt update &&  apt upgrade -y
RUN apt install -y gcc-8 make g++
RUN update-alternatives --install /usr/bin/gcc gcc /usr/bin/gcc-8 2000
WORKDIR /files
COPY ./files .
RUN g++ -O3 ./src/bbfs_node.cpp -o out
RUN g++ -O3 ./src/bbfs_edge.cpp -o out
\end{lstlisting}

Experiment E7 is more complex than E3, as it needs two PLs to execute (Java, and Unix Shell) and two associated dependencies, ``maven'' and ``libgomp1''.
The configuration file of the experiment mentioned that the requirement is ``Java 8''.
To conclude, it is necessary to execute the following command \verb|mvn clean install|. 

In Listing~\ref{lst:dockerfile7}, we can see the dockerfile generated for E7. This file follows the same structure as the previous one, although the concrete commands change according to the experiment's needs.

\begin{lstlisting}[language=json,firstnumber=1, caption={Dockerfile of E7},captionpos=b, label={lst:dockerfile7} ]
FROM ubuntu:20.04
RUN apt update && apt upgrade -y
RUN apt install -y openjdk-11-jdk openjdk-11-jre maven libgomp1
WORKDIR /files
COPY ./files .
RUN mvn clean install
\end{lstlisting}

\subsubsection{Run the Experiments}\label{subsection:runExperiment}
We use the Code Execution component (\cref{sec:code}) to execute the experiment, particularly the endpoint \\\texttt{/\textless projectuuid\textgreater\allowbreak/run-container}, used as parameters \textit{i)} the identifier of the created Docker image in the previous task and \textit{ii)} the command we intend to run.
Listing~\ref{lst:runExperimentE3} presents the request data to execute E3. Listing~\ref{lst:runExperimentE7} presents the request data to execute E7.

\begin{lstlisting}[language=json,firstnumber=1, caption={Request data for the execution of E3},captionpos=b, label={lst:runExperimentE3} ]
{"tagId": "e3", 
 "command": "./out freebase/edges.txt freebase/labels.txt 1234 5678 4 0.3 2000 0.8 1 10 0.01 2"}
\end{lstlisting}

\begin{lstlisting}[language=json,firstnumber=1, caption={Request data for the execution of E7},captionpos=b, label={lst:runExperimentE7} ]
{"tagId": "e7", 
 "command": "/bin/confetti -c $(scripts/examples_classpath.sh) edu.berkeley.cs.jqf.examples.closure.CompilerTest testWithGenerator fuzz-results"}
\end{lstlisting}

\subsubsection{Build the Research Artifact}\label{subsection:reproducible_package}
Finally, we employ the Code Execution component (\cref{sec:code}) to build the research artifact.

We use the container image generated in the building of the computational environment task and reuse all the information provided during the execution of the experiment to encapsulate all that information in a folder along with the scripts needed. For that, we make use of the endpoint \texttt{/\textless project-uuid\textgreater/package} and insert as parameters the commands used to run the experiment (as in \cref{subsection:runExperiment}), as well as the experimental
identifier of the created Docker image (as in \cref{sec:ex_env}).
We built a research artifact that can run on Windows, macOS, or Unix OS where the user just needs to have Docker installed and running. By clicking in the ``runExperiment'' file, the experiment is executed automatically.

We created a research artifact for each of the experiments, which we make publicly available~\cite{lazaro_2024_artifact}.

\subsection{Answering RQ2}
To address RQ2, we developed a framework that supports the creation, execution, and validation of computational experiments across various fields. This framework facilitates the reproducibility and replicability of experiments by providing an API to configure computational environments via containerization technologies, such as Docker. Our framework guides researchers through a structured process to create research artifacts, which include all necessary components such as code, datasets, and execution commands. By supporting multiple PLs and database configurations, our framework ensures broad applicability. The ability to manage experimental files, infer dependencies, and validate results makes our approach versatile and effective in promoting reproducibility in computational research.

%% file: 6-experimental.tex
Our evaluation aims to assess the framework's ability to handle a diverse range of experiments, from simple scripts to those involving databases, multiple PLs, and various research domains. Additionally, we seek to compare our framework with existing alternatives.

\subsection{Study Design}
We started by collecting experiments already published at scientific conferences. To do so, we used two different computer science research communities: software engineering through the conference \textit{IEEE/ACM International Conference on Software Engineering} (ICSE) and databases through the conference \textit{International Conference on Very Large Databases} (VLDB), two very well-known scientific events in their respective areas. 
We analyzed the first 30 articles of VLDB 2021\footnote{\url{http://vldb.org/pvldb/volumes/15/}} and the first 50 articles of ICSE 2022\footnote{\url{https://ieeexplore.ieee.org/xpl/conhome/9793835/proceeding?isnumber=9793541}}.
For each publication, we collected the link where the code of the publication is available. However, many of the publications do not have this information: for ICSE, only 32 papers have a link, and for VDLB, only 19. From those with code, we randomly selected five publications from each conference.

We also tried to find computational experiments in other areas through the Climate Change\footnote{\url{https://www.springer.com/journal/10584/}} and Nature Climate Change\footnote{\url{https://www.nature.com/nclimate/}} journals, but the papers did not include information about the setup. Without this information, we could not reproduce them. 

Furthermore, we have widened the range of experiences, and to do this, we have made three searches in the Zenodo repository, a well-known, open-science repository \cite{zenodo}. We search with ``Medical'', ``Artificial Intelligence'', and ``Climate Change'' expressions. We considered only software repositories and the first 100 results of each repository. Once again, we randomly select five results for each query.

We chose computer science conferences, medical, artificial intelligence, and climate change experiments, expecting that these works would pose more complex challenges and a variety of experiences from various domains.

In addition, we performed all the experiments applicable to all related approaches (see \cref{sec:sota_tools}).
%
For that, we collected all the published use case experiments in those approaches. However, only two published articles contained experiments that were still available and complete.
From the works SciInc~\cite{Youngdahl2019SciInc} and Sciunit~\cite{Ton2017SciUnits} we collected three experiments: the Chicago Food Inspections Evaluation (E26)~\cite{e26}, the Variable Infiltration Capacity (E27)~\cite{e27}, and the Incremental Query Execution (E28)~\cite{e28}. 

We then proceeded to collect all the relevant information to reproduce the experiment. In \cref{tab:dataset}, we present the relevant information related to the set of experiments chosen from VLDB (experiments E1--E5), from ICSE (E6--E10), for climate change (E11--E15), for medicine (E15--E20), for artificial intelligence (E21--E25) and from related work (E26--E28).
For each experiment, we present (in this order) the experiment identifier, the original publication, the PLs used, the database engine used (if any, with \cross indicating that no database is used, a link to the original repository (clickable in the PDF version of the paper), whether the dependencies are described (\mycheckmark indicates that there is a sufficient description, \cross means that there is not a description, and \halfcheckmark\ that the description is not clear), indicates how to execute the experiment (\mycheckmark indicates a sufficient description, \cross indicates insufficient information, and Package is used for repositories that are libraries, packages, or similar artifacts that do not produce a specific result), and finally, the indication if we could execute it via our framework. 
We follow the \textit{Artifact Review and Badging (v1.1)}\footnote{\url{https://www.acm.org/publications/policies/artifact-review-and-badging-current}} guidelines from ACM, applying the following classifications:

\begin{description}
    \item[\textit{Artifacts Evaluated - Functional}]\includegraphics[width=0.4cm]{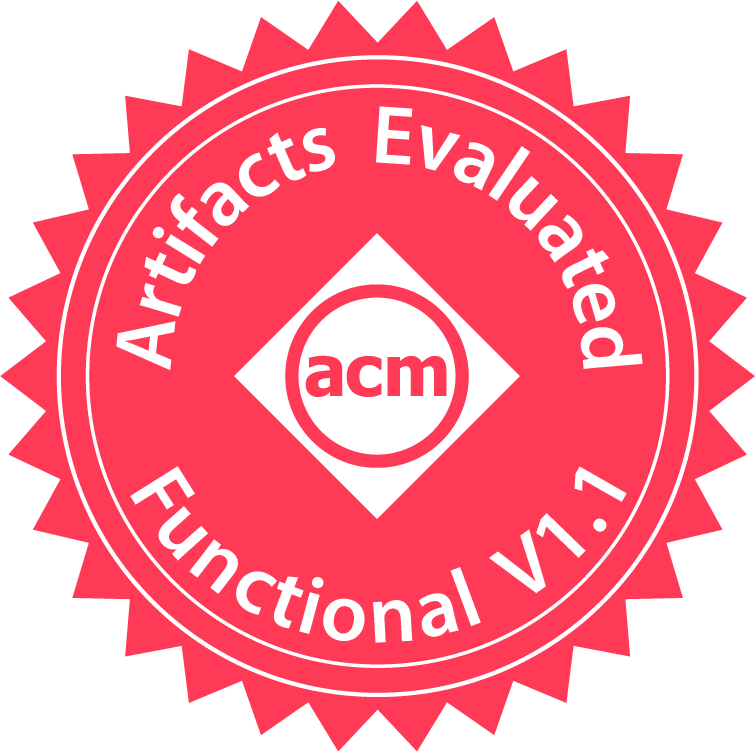} – The artifact has been reviewed and verified to be documented, complete, and functional.
    
   \item[\textit{Artifacts Evaluated - Reusable}] \includegraphics[width=0.4cm]{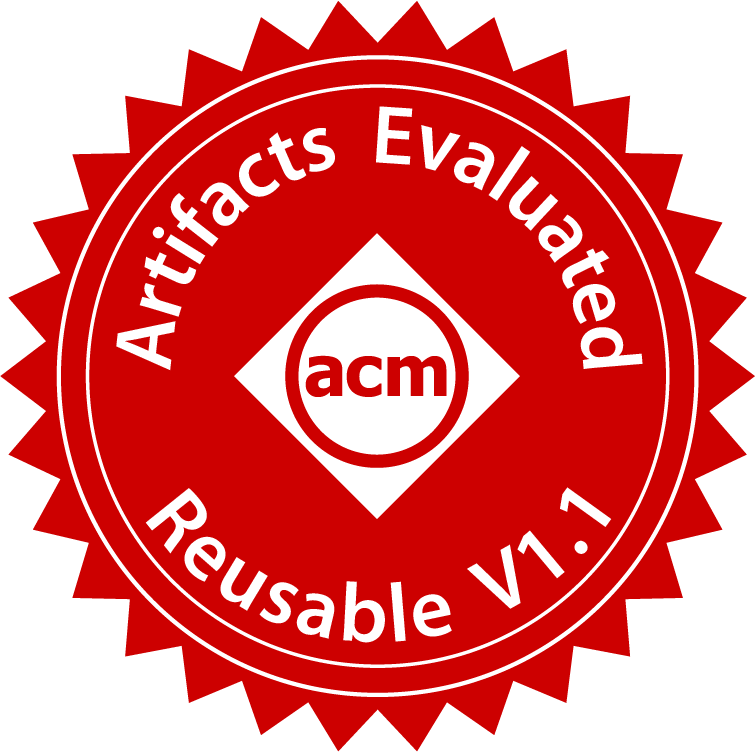}– The artifact provides significant functionality and is well-documented for reuse by other researchers.
    
    \item[\textit{Artifacts Available}]\includegraphics[width=0.4cm]{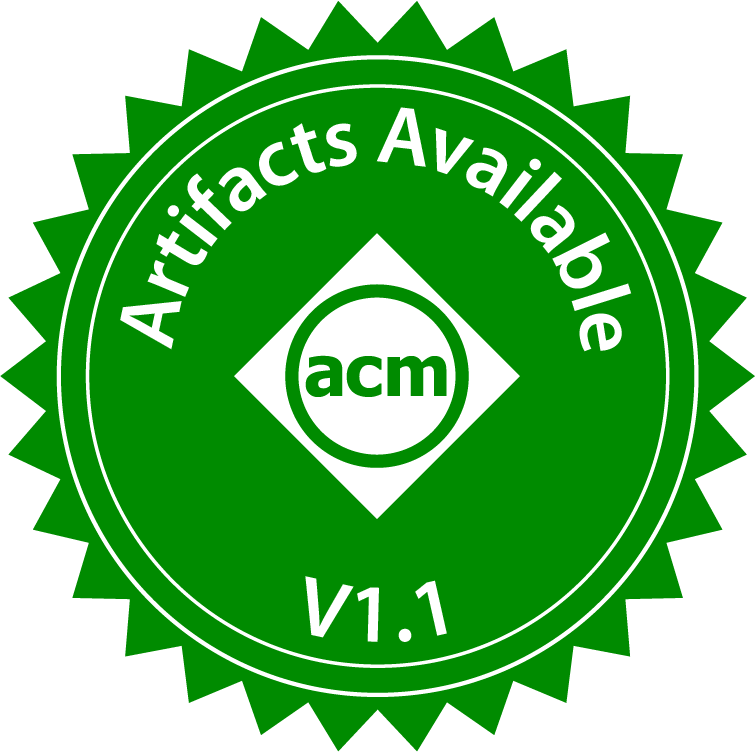} – The artifact is stored in a publicly accessible repository with a unique identifier, ensuring long-term availability.
    
   \item[\textit{Results Validated - Reproduced}]\includegraphics[width=0.4cm]{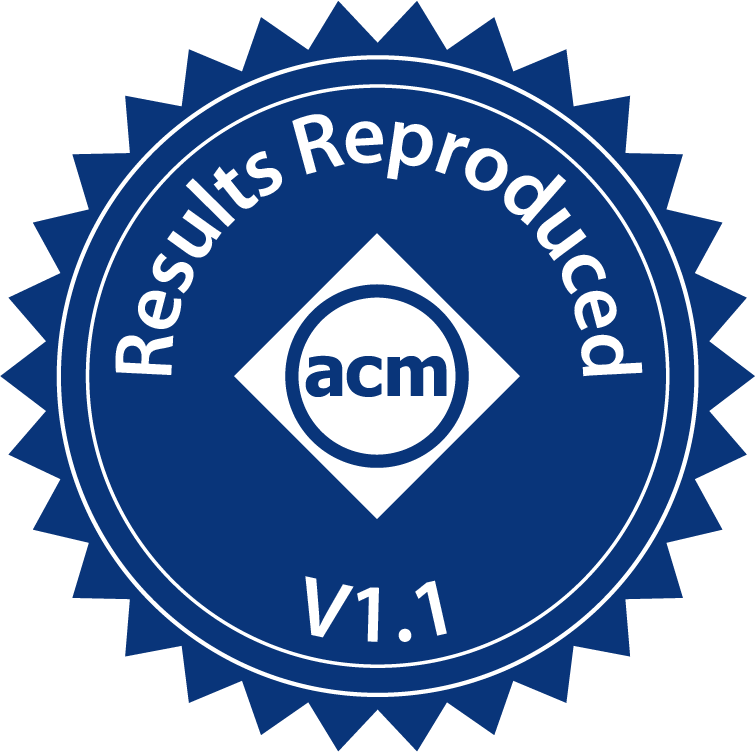} –  This artifact indicates that the main results of the original study have been successfully obtained by an independent person or team using, at least in part, the artifacts provided by the original authors. In our evaluation, we focused on this category by executing the complete source code from each repository to verify reproducibility. We did not consider the \textit{Results Validated - Replicated} badge, as replication requires an independent reimplementation using different methods or conditions.

\end{description}

Additionally, we classify experiments as \cross when the experiments could not be executed, and we provide a footnote explaining the reason for the failure. Not available is used to mark repositories that are inaccessible or no longer exist and NA means not applicable is used to mark repositories that are packages or similar artifacts that do not produce a specific result.


\subsection{Executing the Experiments Using our Framework}
For each experiment, it was necessary to create and add the corresponding code to each project, build the necessary environment, run the experiments, and finally create a research artifact. For each experiment, we followed the procedures of the examples described in \cref{sec:examples}.

Our own experimental environment was executed on a personal computer with a 6-core CPU and 16 GB of RAM running Ubuntu 20.04.

\subsection{Results}
As presented in \cref{tab:dataset}, the dataset of 28 experiments is available in \cite{costa_2024_dataset}. However, four of these experiments are packages or libraries and thus cannot be considered for a reproducibility study, as they are not used to run or produce any results presented in the corresponding papers.
Additionally, six other experiments do not have the source code or data available, making them impossible to execute.
Thus, only 18 experiments are available for reproduction. We were able to successfully install the necessary dependencies and run and create a research artifact for 16 of the 18 (89\%).

To validate computational reproducibility, we compared the outputs from our framework with expected results documented in research articles or available in GitHub/Zenodo repositories. For experiments E13, E19, E26, and E28, we validated the output produced by our framework by comparing it with the expected result available in the original code repository.
For the remaining experiments, we compared the results with those available in scientific publications.
In all the cases, the produced outputs are the ones reported by the experiments. Thus, our framework was able to reproduce 100\% of the experiments it was able to execute.


We could not run E8 because we could not find a Java class during the execution.
For E18, there is no information about how to install the project, so we tried to install the project using \verb+python -m pip install bio2bel_mesh+ command. However, the package requires Python 3.5, which is currently incompatible with several other packages. 
%


Notably, we were not able to run E8 and E18 on our personal computer or on other reproducibility frameworks (Code Ocean, RenkuLab, and Whole Tale).
The problem may be related to a possible error in the source code or some failure in the configuration of the computational environment due to a lack of information on the part of the authors. Hence, it is important to provide the research artifact of the experiment and not just the source code and the necessary dependencies. Thus, our success rate of 89\% is very conservative, as we can argue that we were able to reproduce 100\% of the experiments that provided complete content.

\begin{table*}[h!]
\centering
\rowcolors{2}{gray!25}{white}
  \centering
\caption{Experiments used in the study (NA stands for Not Applicable)}
\begin{threeparttable} 
\begin{tabular}{
m{0.4cm}>{\centering}
m{0.65cm}>{\centering}
m{3.7cm}>{\centering}
m{1.8cm}>{\centering}
m{1.75cm}>{\centering}
m{1.8cm}>{\centering}
m{1.6cm}>{\centering\arraybackslash}
m{1.65cm}}\rowcolor{gray!50}
  
  \toprule
  Id. & Pub. & PLs & Database & Repository link & Dependencies description & How to execute & Reproduced \\
  
  \midrule
    
 E1 & \cite{Adnan2022} & Unix Shell, Python & \cross & \href{https://github.com/STAR-Laboratory/Accelerating-RecSys-Training}{\underline{Link}} & \mycheckmark  & \mycheckmark & \includegraphics[width=0.4cm]{images/E-R.png} \includegraphics[width=0.4cm]{images/reproduced.png}  \\
 E2 & \cite{Bai2022} & Python & \cross & \href{https://github.com/jiyangbai/TaGSim}{\underline{Link}} & \cross & \mycheckmark & \includegraphics[width=0.4cm]{images/E-R.png} \includegraphics[width=0.4cm]{images/reproduced.png} \\
 E3 & \cite{Chauhan2022} & C++
 & \cross & \href{https://github.com/idea-iitd/RQuBE}{\underline{Link}} & \cross & \mycheckmark & \includegraphics[width=0.4cm]{images/E-R.png} \includegraphics[width=0.4cm]{images/reproduced.png}\\
 E4 & \cite{Sun2022}  & Python & PostgreSQL & \href{https://github.com/jt-zhang/CardinalityEstimationTestbed}{\underline{Link}} & \cross & \mycheckmark & \includegraphics[width=0.4cm]{images/E-R.png} \includegraphics[width=0.4cm]{images/reproduced.png}\\
 E5 & \cite{Zhou2022} & C++ & \cross & \href{https://github.com/AlexanderTZhou/IUBFC}{\underline{Link}} & \mycheckmark & \mycheckmark & \includegraphics[width=0.4cm]{images/E-R.png} \includegraphics[width=0.4cm]{images/reproduced.png} \\
   
   \midrule
 E6 & \cite{Chen2022}  & Python & \cross    & \href{https://github.com/OpsPAI/ADSketch}{\underline{Link}} & \mycheckmark & \mycheckmark & \includegraphics[width=0.4cm]{images/E-R.png} \includegraphics[width=0.4cm]{images/reproduced.png}\\
 E7 & \cite{Kukucka2022} & Java, Python, Unix Shell & \cross & \href{https://github.com/neu-se/CONFETTI}{\underline{Link}} & \mycheckmark & \mycheckmark & \includegraphics[width=0.4cm]{images/E-R.png} \includegraphics[width=0.4cm]{images/reproduced.png}\\
 E8 & \cite{Nguyen2022} & Python, Java & \cross & \href{https://github.com/hub-se/BeDivFuzz}{\underline{Link}} & \mycheckmark & \mycheckmark & \cross{}\tnote{a} \\
 E9 & \cite{Xie2022}  & Python & \cross    & \href{https://github.com/ICSE2022FL/ICSE2022FLCode}{\underline{Link}} & \mycheckmark & \mycheckmark  & \includegraphics[width=0.4cm]{images/E-R.png} \includegraphics[width=0.4cm]{images/reproduced.png}\\
 E10 & \cite{Zhang2022} & \multicolumn{6}{c}{Not available}\\
    
    \midrule
 E11 & \cite{edwards2022}  & R & \cross & \href{https://zenodo.org/record/5967578#.ZEwQKnbMJD8}{\underline{Link}} & \mycheckmark &     Package & NA \\
 E12 & \cite{Hemes2023}  & R & \cross & \href{https://zenodo.org/record/7651416#.ZEwPxXbMJD8}{\underline{Link}} & \mycheckmark & \cross    & \cross{}\tnote{c} \\
 E13 & \cite{Jehn2021} & Python & \cross & \href{https://zenodo.org/record/4588383#.ZEwRHnbMJD8}{\underline{Link}} & \cross & \cross & \includegraphics[width=0.4cm]{images/E-R.png}  \includegraphics[width=0.4cm]{images/Available.png} \includegraphics[width=0.4cm]{images/reproduced.png}  \\
 E14 & \cite{Lin2024} & Python & \cross & \href{https://zenodo.org/record/7613549#.ZEwQSnbMJD8}{\underline{Link}} & \cross & \mycheckmark & \includegraphics[width=0.4cm]{images/E-R.png}  \includegraphics[width=0.4cm]{images/Available.png} \includegraphics[width=0.4cm]{images/reproduced.png}\\
 E15 & \cite{Qasmi2021} & R & \cross & \href{https://zenodo.org/record/5233947#.ZFPecHbMJD8}{\underline{Link}} & \mycheckmark & Package & NA  \\

    \midrule
 E16 & \cite{Frie2021} & Python & \cross & \href{https://zenodo.org/record/4497214#.ZFQy33bMJD-}{\underline{Link}} & \cross & \mycheckmark &  \cross{}\tnote{c} \\
 E17 & \cite{Tianyu2021}  & Python & \cross & \href{https://zenodo.org/record/4926118#.ZEFWgnbMJD8}{\underline{Link}} & \mycheckmark & \mycheckmark & \cross{}\tnote{c}  \\
 E18 & \cite{Hoyt2018} & Python & \cross & \href{https://zenodo.org/record/1488650#.ZFQw4nbMJD_}{\underline{Link}} & \mycheckmark & \mycheckmark & \cross{}\tnote{a}  \\
 E19 & \cite{Kailas2022}  &  Python, Unix Shell & \cross & \href{https://zenodo.org/record/6617957#.ZEFTEXbMJD8}{\underline{Link}} & \mycheckmark & \mycheckmark  & \includegraphics[width=0.4cm]{images/E-R.png}  \includegraphics[width=0.4cm]{images/Available.png} \includegraphics[width=0.4cm]{images/reproduced.png} \\
 E20 & \cite{Yang2021} & Python & \cross & \href{https://zenodo.org/record/4453803#.ZFQxW3bMJD-}{\underline{Link}} & \mycheckmark & \mycheckmark & \includegraphics[width=0.4cm]{images/E-R.png}  \includegraphics[width=0.4cm]{images/Available.png} \includegraphics[width=0.4cm]{images/reproduced.png} \\
  \midrule
 E21 & \cite{Prezja2013} & Python & \cross & \href{https://zenodo.org/record/7867100}{\underline{Link}} & \mycheckmark & Package & NA   \\
 E22 & \cite{Santucci2023} & Python & \cross & \href{https://zenodo.org/record/7595510}{\underline{Link}} & \cross & \cross & \includegraphics[width=0.4cm]{images/E-R.png}  \includegraphics[width=0.4cm]{images/Available.png} \includegraphics[width=0.4cm]{images/reproduced.png}\\
 E23 & \cite{Erickson2019} & Python, Unix Shell, C++ & \cross & \href{https://zenodo.org/record/3568468}{\underline{Link}} & \halfcheckmark &  Package & NA  \\
 E24 & \cite{Coupette2022} & Python, Jupyter Notebook & \cross & \href{https://zenodo.org/record/6468193}{\underline{Link}} & \mycheckmark & \cross & \cross{}\tnote{c}  \\
 E25    & \cite{doe2023axom} & Python & \cross & \href{https://zenodo.org/record/7585845}{\underline{Link}} & \mycheckmark & \cross & \includegraphics[width=0.4cm]{images/E-R.png}  \includegraphics[width=0.4cm]{images/Available.png} \includegraphics[width=0.4cm]{images/reproduced.png} \\

    \midrule
 E26         & \cite{e26}  & R & \cross & \href{https://chicago.github.io/food-inspections-evaluation/}{\underline{Link}} & \mycheckmark & \mycheckmark & \includegraphics[width=0.4cm]{images/E-R.png} \includegraphics[width=0.4cm]{images/reproduced.png}\\
 E27    & \cite{e27}    & Python, Unix Shell & \cross & \href{https://github.com/uva-hydroinformatics/VIC_Pre-Processing_Rules}{\underline{Link}} & \cross & \cross  & \cross{}\tnote{d} \\
 E28& \cite{e28}   & Python & SQlite & \href{https://bitbucket.org/TonHai/iqe/src/master/}{\underline{Link}} & \cross & \cross & \includegraphics[width=0.4cm]{images/E-R.png} \includegraphics[width=0.4cm]{images/reproduced.png}\\
       \bottomrule
\end{tabular}
\begin{tablenotes}
\item[a] Code crashed
\item[b] Not Applicable – Unit tests were not available in the repository for package cases.
\item[c] No Data – Some required data files are missing, preventing execution.  
\item[d] Missing Files – Certain essential files were not available in the provided repository.  

\end{tablenotes}
\end{threeparttable}
\label{tab:dataset}
\end{table*}

\begin{table}[h!]
\centering
\rowcolors{2}{gray!25}{white}
  \centering
\caption{Comparison of the effectiveness of reproducibility frameworks with our approach (NA means Not Applicable)}
\begin{threeparttable} 
\begin{tabular}{
m{0.4cm}>{\centering}
m{1.4cm}>{\centering}
m{1.4cm}>{\centering}
m{1.4cm}>{\centering\arraybackslash}
m{1.4cm}}\rowcolor{gray!50}
\toprule
   Id.  & Code Ocean & RenkuLab & Whole Tale & Our  \\
   \midrule
      E1 & \mycheckmark & \mycheckmark & \cross{}\tnote{b} & \mycheckmark  \\
      E2 & \mycheckmark & \mycheckmark & \cross{}\tnote{b} & \mycheckmark  \\
      E3 & \cross{}\tnote{c} & \cross{}\tnote{b} & \cross{}\tnote{b} & \mycheckmark   \\
      E4 & \cross{}\tnote{d} & \cross{}\tnote{d} & \cross{}\tnote{b} & \mycheckmark   \\
      E5 & \mycheckmark & \cross{}\tnote{b} & \cross{}\tnote{b} & \mycheckmark   \\
  \midrule
      E6 & \mycheckmark & \cross{}\tnote{e}  & \cross{}\tnote{b} & \mycheckmark   \\
      E7 & \cross{}\tnote{a} & \cross{}\tnote{b} & \cross{}\tnote{b} & \mycheckmark   \\
      E8 & \cross{}\tnote{a} & \cross{}\tnote{a} & \cross{}\tnote{b} & \cross{}\tnote{a}   \\
      E9 & \mycheckmark & \mycheckmark & \cross{}\tnote{b} & \mycheckmark   \\
      E10  & \multicolumn{4}{c}{Not available} \\
  \midrule
      E11 & \multicolumn{4}{c}{NA} \\
      E12 & \multicolumn{4}{c}{No data} \\
      E13 & \mycheckmark & \mycheckmark &  \cross{}\tnote{b} & \mycheckmark   \\
      E14 & \mycheckmark & \mycheckmark &   \cross{}\tnote{b} & \mycheckmark   \\
      E15 & \multicolumn{4}{c}{NA} \\
  \midrule
      E16 & \multicolumn{4}{c}{No data} \\
      E17 & \multicolumn{4}{c}{No data} \\
      E18 & \cross{}\tnote{a} & \cross{}\tnote{a} & \cross{}\tnote{b} & \cross{}\tnote{a}   \\
      E19 & \cross{}\tnote{e} & \cross{}\tnote{e} & \cross{}\tnote{b} & \mycheckmark   \\
      E20 & \mycheckmark & \mycheckmark   &  \cross{}\tnote{b} & \mycheckmark   \\
  \midrule
      E21 & \multicolumn{4}{c}{NA} \\
      E22 & \mycheckmark & \mycheckmark &   \cross{}\tnote{b} & \mycheckmark  \\
      E23 & \multicolumn{4}{c}{NA} \\
      E24 & \multicolumn{4}{c}{No data} \\
      E25 & \mycheckmark & \mycheckmark &  \cross{}\tnote{b} & \mycheckmark  \\
  \midrule
      E26 & \mycheckmark & \mycheckmark &  \mycheckmark & \mycheckmark  \\
      E27 & \multicolumn{4}{c}{Miss files} \\
      E28 & \cross{}\tnote{e} & \cross{}\tnote{e} &         \cross{}\tnote{b} & \mycheckmark  \\
       \bottomrule
\end{tabular}
\begin{tablenotes}
\item[a] Code crashed
\item[b] PL not supported
\item[c] g++ compiler not supported
\item[d] Database not supported
\item[e] Python version not supported
\end{tablenotes}
\end{threeparttable}
\label{tab:comparison}
\end{table}

\subsection{Comparison with Existing Reproducibility Frameworks}
In this section, we present a comparison of the execution of our framework against other existing ones.
In previous work, we evaluated the efficiency of tools designed to assist researchers from various fields in addressing reproducibility challenges. We conclude that only Code Ocean, RenkuLab, and Whole Tale were able to reproduce 100\% of the experiments used~\cite{costa2024Rep}. 
%
%
Thus, for these three frameworks and ours, we executed all the possible experiments from the 28 we presented before and checked the consistency of the results obtained with those provided in the original experimental work.

In \cref{tab:comparison}, we present the results, and for each non-executed experiment, we expose in a footnote the reason that reproduction was impossible. All the information about the experiments executed in Code Ocean, Renkulab, and Whole Tale can be found in our reproducibility package~\cite{lazaro_2024_artifact}. 

The two experiments we could not reproduce in SciRep could not be reproduced by the other tools. The Code Ocean was successful in 11 of the 18 experiments (61\%),  RenkuLab was successful in 9 of the 18 experiments (50\%), Whole Tale was able to execute only 1 (6\%), and our framework was able to reproduce 16 experiments (89\%). Thus, our framework is the most successful, as it can reproduce all experiments containing the necessary information and artifacts and has no code issues.

\subsection{Answering RQ3}
In addressing RQ3 and reinforcing RQ2, we conducted an extensive evaluation of our framework's capabilities by replicating experiments sourced from prestigious conferences, such as ICSE and VLDB, as well as from fields including medicine, artificial intelligence, and climate change. Our objective was to validate the ability of the framework to faithfully reproduce a diverse set of experiments.

Across the selected experiments, which ranged from simple scripts to those integrating databases and complex algorithms, SciRep demonstrated robustness in recreating exact environments as it was able to reproduce 100\% of the experiments it could execute. 

Compared with the state of the art, it is clear that our framework is clearly more effective, as it was able to execute most experiments (89\%), whereas the other tools were successful for only 61\% of the experiments.

Thus, our framework demonstrated the ability to support and accurately reproduce experiments across diverse scientific domains, highlighting its potential to facilitate rigorous and transparent research practices.

%% file: 7-Discussion.tex
In this work, we propose a new framework to overcome the current shortcomings of existing reproducibility tools \cite{costa2024Rep}. Its main goal is to address a wide range of experiments and scientific fields.

We evaluated the capabilities of SciRep by assessing whether it could reproduce the results from a variety of experiments across different scientific fields. For this purpose, we randomly chose 28 different experiments from published scientific work in two research communities (ICSE and VLDB), three research areas (medical, artificial intelligence, and climate change), and experiments executed via related approaches. 

Although 28 may not seem to be a significant number, performing an experiment designed and implemented by others is always quite challenging. Moreover, this is even more the case due to the diversity of experiments and domains considered, which encompass different areas and a variety of computational procedures.
During the configuration of the computational environments, we needed to infer the information from the repositories (\eg, PLs and dependencies used) to try to fill in the parameters that were necessary to insert to build the same computational environment.

Nevertheless, the experiments ran smoothly. Validating reproducibility ranged from comparing the contents of the terminal output to checking and analyzing the contents of the output files. Although the repositories included the command lines to execute the code and the expected results, checking the reproducibility and consistency of the results obtained was challenging. Our framework checks the differences between two executions, but we had to ensure that the results obtained were truly the correct ones, requiring some manual work in this verification. However, after this first analysis, it is possible to mark one execution's result as correct. From then on, all runs are compared with this result.



In addition, our approach successfully reproduced more experiments than the three state-of-the-art reproducibility tools did. Indeed, even professional tools, such as Code Ocean, could reproduce only approximately half of the experiments, highlighting the need for our work.

There were, however, two experiments we could not execute, as we cannot run such experiments with any of the other three tools or on our computers. This indicates that there are fundamental problems with the experiments and their code or data that do not allow others to run them. 

Our tool also addresses the three challenges described in \cref{sec:challengesInReproducibility}. 

Software (Challenge 1) is the biggest problem when addressing reproducibility. Our proposal allows users to build a complete container of the project, which will make it available for reuse by other researchers, even when some dependencies are no longer publicly available, as the artifact acts as a capsule. This addresses principles P1 and P2. 

The Command (Challenge 2), which is implemented in our approach, allows researchers to configure the execution of the experiments and their execution with the same parameters. The results of previous executions can be used as a comparison to validate the computational reproducibility. Principles P3 and P4 are addressed in this part of the work. 

Data (Challenge 3) is handled in the framework by providing a mechanism to allow researchers to manage the data of the experiment. This is principle P5. Nevertheless, in this work, we do not consider the challenges of big data or even the privacy that some data require. These are interesting future directions for our research.

Our approach is able to handle diverse experiments, from the execution of simple scripts to the execution of complex experiments (\eg, involving databases or the simultaneous use of several PLs). As a result, our proposal effectively addresses principles P6 and P7.

Moreover, the research artifact that can be obtained from the framework can be easily shared, as it corresponds to a compressed file with just two files.
By executing one, with a simple double-click, the entire experiment is executed. The only requirement is that the Docker software is installed and running. Thus, we have reduced the problem of re-execution to having to install a single software tool, which is the most commonly used software of its kind and is thus expected to endure \cite{dockeribm, dockeroracle, Amit2020}. Thus, principles P9 and P10 are also addressed.

\subsection{Limitations}

A common challenge in container-based approaches, such as ours, is the possible future need to apply updates, fix bugs, or extend replicability over time. While our framework currently does not provide a simplified method for modifying code or correcting errors after the research artifact has been generated, researchers can still manually execute the container associated with the created image and apply the necessary changes.
However, such modifications require technical expertise, as they involve interacting with the containerized environment, potentially rebuilding the image, and ensuring that any adjustments maintain the integrity of the original computational experiment. This limitation highlights the trade-off between strict reproducibility and practical flexibility in research workflows. Future improvements to the framework may explore more accessible mechanisms for controlled modifications, balancing replicability with adaptability to evolving research needs.
Additionally, tools like Nix\footnote{\url{https://nixos.org/}} and Guix\footnote{\url{https://guix.gnu.org/}} offer advanced dependency management, ensuring consistent and isolated environments across different systems. However, while they facilitate software dependency and configuration management, they do not inherently ensure computational reproducibility and primarily focus on Linux-based systems.

Another common concern in container-based reproducibility is the dependency on container images such as Docker Hub. While relying on Docker Hub poses potential risks, such as service disruptions or changes in availability, it remains widely used by industry and research institutions due to its convenience and extensive ecosystem. 

The size of the resulting images also poses some challenges in storing and sharing the experiments.
Nevertheless, while container images can become large, they are generally smaller than VMs, which are the alternative recommended by ACM for research artifacts\cite{zhang2018}. 

While the current implementation effectively automates the setup of environments for reproducibility, the concrete code could be further improved. In particular, the inference mechanism for PLs and their dependencies can only be expanded to new languages by updating the source code. Future work will focus on making such changes more user-friendly by devising abstractions representing PLs and ways to detect them so SciRep can cope with more languages without the need to update its code. For instance, this information could be saved in a database or even configuration files that would then be processed by the tool.

Our approach is a framework for reproducibility, but it does not provide a user interface. It is not intended to be used by researchers directly at this time; however, we have validated the reproducibility potential of the framework, and we plan to design and build such an interface in future work~\cite{costa2022platform,costa2024vlhccGC}.
Thus, principle P8, which we consider very important, is not yet supported by our work.

%% file: 8-threats.tex
The validation process has some threats~\cite{Sarah2010, Cook1979, Wohlin2012}, which we address next.



A possible threat to the generalizability of our results relates to the experiments used.
However, we collected a variety and a significant number of experiments (\eg, with only one PL, with more than one PL, and with and without the use of databases). Moreover, we selected experiments from different scientific fields, including climate change, artificial intelligence, software engineering, databases, and medicine. 
We were successful in reproducing all the experiments that had enough information and artifacts to be executed. 
This provides significant evidence that SciRep can indeed handle a variety of scientific experiments.

In most cases, the computational environment is only briefly described by the authors of the experiments we used, so it is more difficult to infer which dependencies and which versions were used in the execution of the original experiment. Thus, there is a risk that we have executed the experiments under slightly different conditions. Nevertheless, we were able to validate the results we obtained with those in the original publications; thus, this risk is quite low.

We compared our approach with three others. 
In these cases, of self-evaluation, there was always a risk of bias.
However, the reasons for the lack of execution of the other tools are quite clear and factual, and we make all the material available for verification \cite{lazaro_2024_artifact}.


%% file: 9-conclusion.tex
One of the greatest obstacles to achieving reproducible computational research is creating the same computational environment, configuring it correctly, reproducing the experiment, and verifying its reproducibility.

Thus, we propose SciRep, a framework designed to support research in multiple fields and ensure that the research is reproducible, replicable, and transparent. Our goal is to have a framework that can reproduce different kinds of experiments, from the execution of simple scripts to more complex multiple PLs experiments.

To evaluate our approach, we collected a set of experiments from different scientific fields with varying complexity. The experiments we used are notably complex in terms of the number of PLs, databases, and other factors. This diversity and complexity provide empirical evidence that our framework can effectively support experiments from multiple domains. By addressing the challenges of reproducibility and replicability across different scientific fields, SciRep makes a significant contribution to the current landscape of computational research.

Our evaluation of the approach revealed that we successfully recreated 16 out of 18 experiments (89\%), whereas current approaches support only approximately 61\%. Moreover, SciRep reproduced 100\% of the executed experiments, meaning that it was able to faithfully recreate an environment that allowed it to run the experiments and achieve the original results. This demonstrates the need for our framework. Moreover, the two experiments we could not reproduce were also not reproducible in the other frameworks, which supports the fact that the problem was with the experiments and not with our tool. 

Finally, it is important to note that SciRep is the only one providing an API, that can be used by others to build their own reproducibility environment, possibly tailored for specific domains.

In future work, we will devise a user interface to enable researchers to create research artifacts with ease~\cite{costa2022platform,costa2024vlhccGC}. Furthermore, we intend to analyze the features of user study tools identified in previous work and design ideas that might be adapted to reproducibility tools~\cite{costa2024vlhcc}.

%% file: appendixApi.tex
\subsection{Create a new Project}
\textbf{Endpoint}: \texttt{POST /project} \\
A new project with a specified name and optional description is created.

\textbf{Request Body}: \texttt{name} (string, required) the name of the project; \texttt{description} (string, optional) a description of the project.

\textbf{Response}: \texttt{projectUuid} (string) is the unique identifier (UUID) of the created project.

\subsection{Upload a file to a Project}
\textbf{Endpoint}: \texttt{POST /:projectUuid/uploadfile} \\
A file is uploaded to the specified project.

\textbf{Parameters}: \texttt{projectUuid} (string, URL parameter) is the unique identifier (UUID) of the project.

\textbf{Request Body}: \texttt{parentDirectory} (JSON object) the directory where the file will be uploaded; \texttt{fileInformation} (JSON object) details the file being uploaded.

\textbf{Form Data}: \texttt{file} (file) the file to be uploaded.

\subsection{Upload a Project}
\textbf{Endpoint}: \texttt{POST /:projectUuid/uploadproject} \\
Uploads an entire project, including all its files and directories.

\textbf{Parameters}: \texttt{projectUuid} (string, URL parameter) is the unique identifier (UUID) of the project.

\textbf{Request Body}: project files to be uploaded.

\subsection{Upload a Project from a remote repository}
\textbf{Endpoint}: \texttt{POST /:projectUuid/uploadgitproject} \\
Uploads a project from a remote repository.

\textbf{Parameters}: \texttt{projectUuid} (string, URL parameter) is the unique identifier (UUID) of the project.

\textbf{Request Body}: \texttt{repositorySelected} (string) the type of the repository (e.g., GitHub, Figshare); \texttt{projectLocation} (string) is the URL or DOI of the remote project.

\subsection{Build Docker Image}
\textbf{URL}: \texttt{POST /<projectUuid>/build-docker-image} \\
A Docker image is built on the basis of the provided configurations and optional database settings.

\textbf{Request Body}: \texttt{configurationForm} includes a Dockerfile, database Dockerfile, and configuration details; \texttt{projectUuid} is the unique identifier for the project; \texttt{DBhas, dbConfiguration} (optional) fields for database configuration.

\textbf{Response}: Returns the Docker image ID.

\subsection{Run Docker Container}
\textbf{URL}: \texttt{POST /<projectUuid>/run-container} \\
Runs a Docker container using the specified image and command. Supports running containers with database configurations.

\textbf{Request Body}: \texttt{command} the command to execute within the container; \texttt{tagId} the Docker image tag ID; \texttt{Optional} configuration form details for databases.

\textbf{Response}: Returns the container logs.

\subsection{Package Research Artifact}
\textbf{URL}: \texttt{GET /<projectUuid>/package} \\
Packages project files and Docker images into a zip file for research artifact purposes.

\textbf{Request Body}: Contains information about commands to run, Docker tag ID, and optional database configurations.

\textbf{Response}: Returns a downloadable zip file containing all the necessary files and scripts.